\documentclass[preprint2]{aastex61}

\usepackage[english]{babel}
\usepackage[utf8x]{inputenc}
\usepackage[T1]{fontenc}
\usepackage{aas_macros}

\usepackage{graphicx} 
\usepackage{amssymb}  %

\begin{document}

\title{Disks around young planetary-mass objects:\\ Ultradeep Spitzer imaging of NGC1333}

\email{as110@st-andrews.ac.uk}

\author[0000-0001-8993-5053]{Aleks Scholz}
\affiliation{School of Physics \& Astronomy, University of St Andrews, North Haugh, St Andrews, KY16 9SS, United Kingdom}

\author[0000-0002-7989-2595]{Koraljka Muzic}
\affiliation{CENTRA, Faculdade de Ci\^{e}ncias, Universidade de Lisboa, Ed. C8, Campo Grande, P-1749-016 Lisboa, Portugal}
\affiliation{Faculdade de Engenharia, Universidade do Porto, Rua Dr. Roberto Frias, 4200-465 Porto, Portugal}

\author[0000-0001-5349-6853]{Ray Jayawardhana}
\affiliation{Department of Astronomy, Cornell University, Ithaca, NY 14853, USA}

\author{Victor Almendros-Abad}
\affiliation{CENTRA, Faculdade de Ci\^{e}ncias, Universidade de Lisboa, Ed. C8, Campo Grande, 1749-016 Lisboa, Portugal}

\author{Isaac Wilson}
\affiliation{Palomar Observatory, California Institute of Technology, Pasadena, CA, United States}

\begin{abstract}
We report on a sensitive infrared search for disks around isolated young planetary-mass objects (PMOs) in the NGC1333 cluster, by stacking 70 Spitzer/IRAC frames at 3.6 and 4.5$\,\mu m$. Our co-added images go $>2.3$\,mag deeper than single-epoch frames, and cover 50 brown dwarfs, 15 of which have M9 or later spectral types. Spectral types $>$M9 correspond to masses in the giant planet domain, i.e., near or below the Deuterium-burning limit of 0.015$\,M_{\odot}$. Five of the 12 PMOs show definitive evidence of excess, implying a disk fraction of 42\%, albeit with a large statistical uncertainty given the small sample. Comparing with measurements for higher-mass objects, the disk fraction does not decline substantially with decreasing mass in the sub-stellar domain, consistent with previous findings. Thus, free-floating PMOs have the potential to form their own miniature planetary systems. We note that only one of the six lowest-mass objects in NGC1333, with spectral type L0 or later, has a confirmed disk. Reviewing the literature, we find that the lowest mass free-floating objects with firm disk detections have masses $\sim0.01\,M_{\odot}$ (or $\sim 10\,M_{\mathrm{Jup}}$). It is not clear yet whether even lower mass objects harbor disks. If not, it may indicate that $\sim 10\,M_{\mathrm{Jup}}$ is the lower mass limit for objects that form like stars. Our disk detection experiment on deep Spitzer images paves the way for studies with JWST at longer wavelengths and higher sensitivity, which will further explore disk prevalence and formation of free-floating PMOs.
\end{abstract}

\section{Introduction}

Planets form in dusty disks surrounding newly born stars. The most straightforward way to identify the presence of circumstellar disks is through infrared imaging, probing for the excess emission from warm dust. The material in the disk is a remnant from the rotating cloud core out of which the star formed, flattened into a disk-shaped structure during the collapse, as angular momentum is conserved. In star forming regions with ages of 1-2\,Myr, more than half of all GKM stars harbor disks. In somewhat older regions with ages of 5-10\,Myr the disk fraction declines to 10-20\%, as the disk material is either accreted onto the stars, blown away by winds, or incorporated into planetary systems \citep{jayawardhana99,haisch01,meyer07}. 

Disks are not only found around stars, but also for young brown dwarfs, substellar objects with masses below 0.08$\,M_{\odot}$, or $80\,M_{\mathrm{Jup}}$, and thus unable to sustain stable Hydrogen burning. Initial discoveries and studies of such circum-sub-stellar disks started about two decades ago \citep{natta02,jayawardhana03,mohanty04}, giving rise to the now well-established fact that most brown dwarfs form like stars from the collapse of cloud cores \citep{luhman12b}. The Spitzer Space Telescope with its unprecendented sensitivity at 3-24\,$\mu m$ was instrumental in furthering our understanding of brown dwarf disks. With growing samples, it became clear that brown dwarf disks are common and long-lived. The disk fractions among brown dwarfs in young star forming regions are comparable to those of coeval stars, translating into similar disk lifetimes \citep{scholz08,luhman12}. Using Spitzer, we also learned that brown dwarf disks show evidence for the growth of dust grains \citep{apai05} and dust settling to the midplane \citep{scholz07} -- prerequisites for planet formation via core accretion. Indeed, subsequent work at longer wavelengths, in particular with ALMA, has shown the potential for planet formation in brown dwarf disks \citep{testi16}. The discovery of several planetary-mass companions orbiting brown dwarfs suggests that substellar objects with masses of 1-8\% the mass of the Sun can form their own planetary systems \citep{jung18}. 

In parallel to the exploration of substellar disks, deep ground-based observing programs have established that star forming regions host  objects with masses that are an order of magnitude below the Hydrogen burning limit. In all regions studied to sufficient depth -- such as, $\sigma$\,Orionis, $\rho$\,Ophiuchi, NGC1333, IC348, Chamaeleon-I, Taurus, Lupus, Upper Scorpius -- surveys find objects with masses below the Deuterium burning limit of $\sim$15$\,M_{\mathrm{Jup}}$ down to masses as low as $\sim 5\,M_{\mathrm{Jup}}$ \citep[e.g.,][]{zapatero00,lucas01,scholz12b,lodieu18,miretroig22}

\begin{figure*}[t]
\center
\includegraphics[width=2.2\columnwidth]{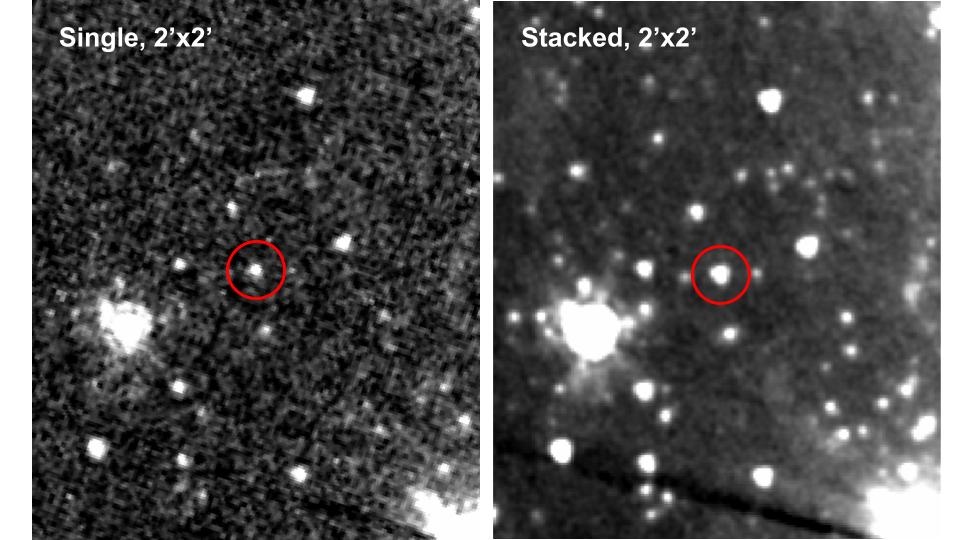}
\caption{Comparison of a single epoch Spitzer/IRAC image of NGC1333 in channel 2 (4.5\,$\mu m$, AOR 29324032) and the stacked version of the same area (70 frames). The marked object in the centre is the L3 brown dwarf SONYC-NGC1333-36, one of the lowest mass spectroscopically confirmed objects in this cluster. The image size is $2'\times 2'$, north is up and east is left.
\label{singlestacked}}
\end{figure*}

We know very little about the nature and evolution of these free-floating planetary-mass objects (PMOs). One basic question relates to their origin: They could be formed either like stars and brown dwarfs, from the collapse of cloud cores \citep{bate12}. They could also be giant planets that got ejected from their natal systems \citep{parker12,vanelteren19}. From purely theoretical arguments, we expect a mix of formation scenarios in the 1-15$\,M_{\mathrm{Jup}}$ mass domain, with a star-like origin more likely at the upper end of the mass range, and a planetary origin dominant at the lower  \citep{scholz22}. The low mass limit for star formation, set by the opacity limit of fragmentation, is expected to be in that mass range, as well \citep{bate12}. But these notions have not been verified empirically yet. Another fundamental question is whether or not these PMOs can form their own miniature planetary systems.\footnote{We continue to call the potential satellites of PMOs 'planets', keeping in mind that if they exist, they would not be orbiting a star, but an object itself comparable to giant planets in mass and size.}

Searching for the signature of disks around free-floating PMOs will shed light on these questions. If PMOs frequently host disks, they have at least the potential to form their own planetary systems. Follow-up studies can then clarify to what extent the prerequisites for planet formation (grain growth, sufficient dust mass, longevity) are met. If free-floating PMOs form predominantly like planets in disks around stars, we would expect their primordial disks to be disrupted \citep{bowler11}. Thus, at face value, we do not expect substantial disks to be common around PMOs if they themselves form in circumstellar disks. Since we expect ejected planets to become more common with decreasing mass, an important benchmark is to determine the mass below which disks become rare or non-existent.

In this paper we present new measurements of the infrared magnitudes of PMOs in the young cluster NGC1333, a compact active star-forming region with an age of $\sim$1\,Myr \citep{gutermuth08,scholz09} and a distance of 300\,pc, according to Gaia DR2 parallaxes of member stars \citep{pavlidou22}. In contrast to previous infrared surveys in this cluster, we construct new, ultradeep Spitzer images by stacking a time-series observation. In Section \ref{data} we present the methodology and the sample. In Section \ref{disc} we derive the infrared excess, verify the presence of disks in PMOs, and discuss PMO disks in general. We present conclusions in Section \ref{sum}. 

\section{Data analysis}
\label{data}

\subsection{Image stacking}

NGC1333 was observed as part of the YSOVAR program during Spitzer's Warm Mission, under program id 61026. The field was observed in IRAC channels 1 ($3.6\,\mu m$) and 2 ($4.5\,\mu m$), in the following called IRAC1 and IRAC2. For full information on YSOVAR, see \citet{rebull14}. In short, YSOVAR was a large-scale monitoring program with Spitzer to study variability in young stars, focused in particular on inner disk structure, accretion changes, eclipsing binaries and rotation periods. The program surveyed 12 star forming regions. The number of epochs varies from region to region, from a minimum of about a dozen to several thousand. NGC1333 was one of the target regions for YSOVAR, with a survey area of $2\times2$ IRAC field of views, where a $10'\times 10'$ field was covered by both channels. For more information on the variability analysis of the NGC1333 dataset, see \citet{rebull15}. The fundamental reference for the IRAC instrument is \citet{fazio04}.

We downloaded 70 images from the YSOVAR dataset on NGC1333, for each of the two channels, from the Spitzer Heritage Archive. The images for a given channel all have very similar pointing, with minimal field rotation, and similar depth. However, due to the layout of the Spitzer focal plane, the images for channel 1 are offset compared to channel 2. Stacking was performed using the Python {\tt reproject} package \citep{robitaille20}, with the function {\tt reproject\_and\_coadd}. We stacked in two iterations, for each filter separately: First we created 7 stacks from 10 images each, then we stacked the resulting 7 stacks to a final deep image for each band. In Figure \ref{singlestacked} we show a comparison for a part of the stacked image vs. the same part of a single image.

\subsection{Sample selection}

We created a catalogue of brown dwarfs in NGC1333 based on the recent census by \citet{luhman16}, which builds on several previous surveys and includes a comprehensive literature review. Their published list comprises 203 new members, the clear majority of which are in the area covered by our deep Spitzer images. Most members have optical or near-infrared spectroscopic confirmation, either by the authors of that census, or from the literature. In addition, the authors use their own astrometry, as well as indicators of youth (e.g., X-ray emission, infrared excess) to identify cluster members. The Luhman census is 'nearly complete' down to K magnitudes of 16.2 and $A_J$ of 3.

From that list, we selected the 65 for which the spectral type adopted by \citet{luhman16} is M6 or later. A spectral type of M6 is commonly used as the boundary between stars and brown dwarfs in star-forming regions. It translates to a temperature of $\sim$3000\,K \citep{muzic14}, corresponding to an object with a mass at or slightly above the substellar limit of 0.08$\,M_{\odot}$ for an age of 1-5\,Myr, according to evolutionary tracks \citep{baraffe15}. That means our sample should safely encompass all known brown dwarfs in the NGC1333 census, but may include some very low mass stars as well. The sample also includes 19 objects with a spectral type of M9 or later, i.e., with estimated masses near or below the Deuterium-burning limit. Those PMOs are the main focus of this paper.

Of the 65 likely brown dwarfs, 49 -- including 14 PMOs -- are covered by at least one of our deep Spitzer/IRAC images. After the \citet{luhman16} census, two more papers confirmed new substellar members for NGC1333: \citet{esplin17b} and \citet{allers20}. The former includes two more objects with M9 or later spectral type, one of which is covered by our images, and was originally reported by \citet{oasa08}. We added it to our sample (with spectral type M9-L4, we adopt L1.5 here). Thus our total sample of PMOs is comprised of 15 objects. Due to the spatial offsets between the images in the two Spitzer/IRAC channels, a few objects in our sample are only covered by one channel. A dozen of the sources are in areas with highly variable background or very close to the edge. Of the 65 likely brown dwarfs, 34 are in the JKS catalogue published by \citet{scholz12b}, which combines the deep J- and K-band photometry from \citet{scholz09} with the list of YSOs created from the the single-epoch Spitzer/IRAC images taken by the C2D project \citep{evans09}. For the objects covered by our deep Spitzer images, 23 (IRAC1) and 22 (IRAC2) are in the JKS catalogue.

\begin{figure*}[t]
\center
\includegraphics[width=1.0\columnwidth]{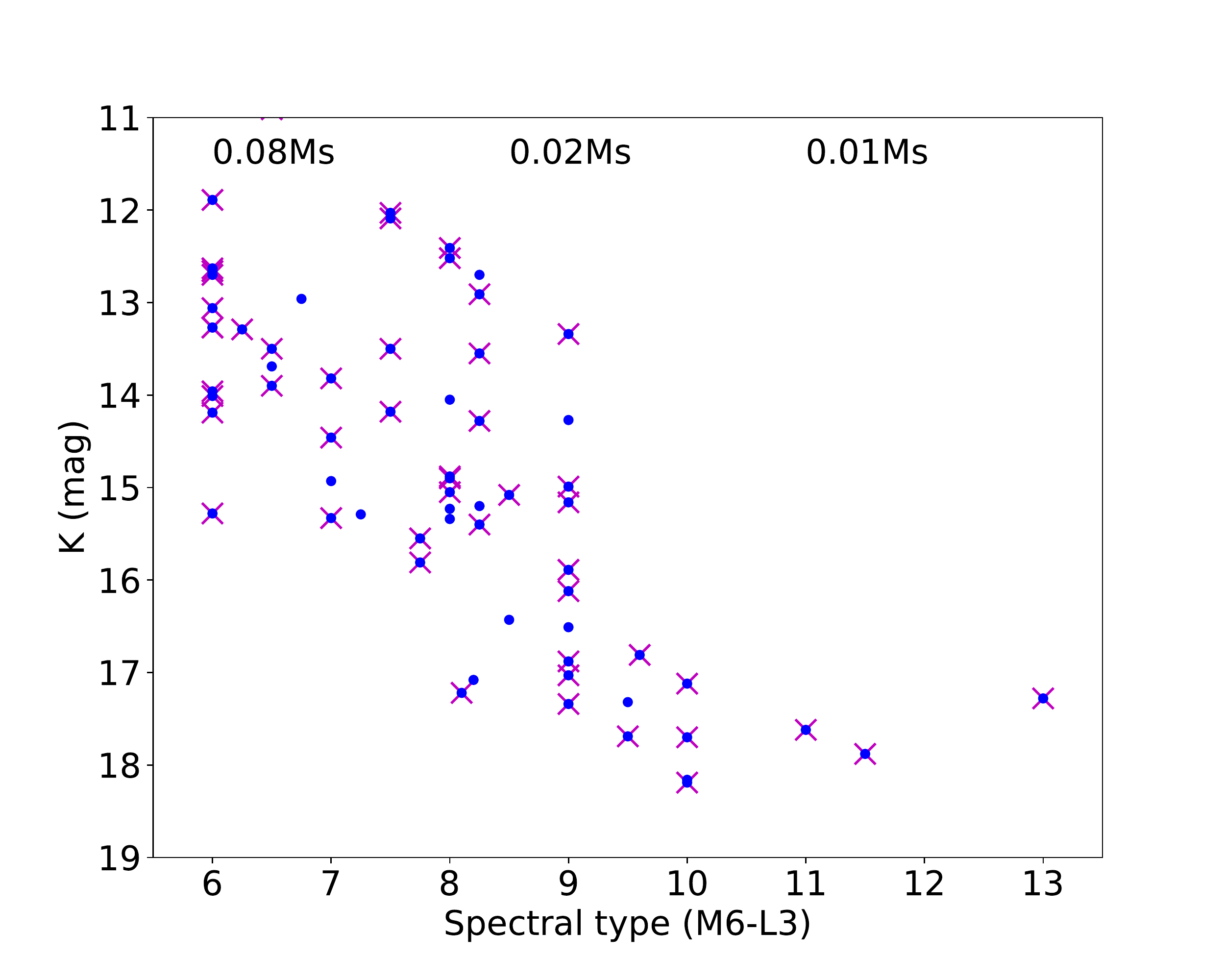}
\includegraphics[width=1.0\columnwidth]{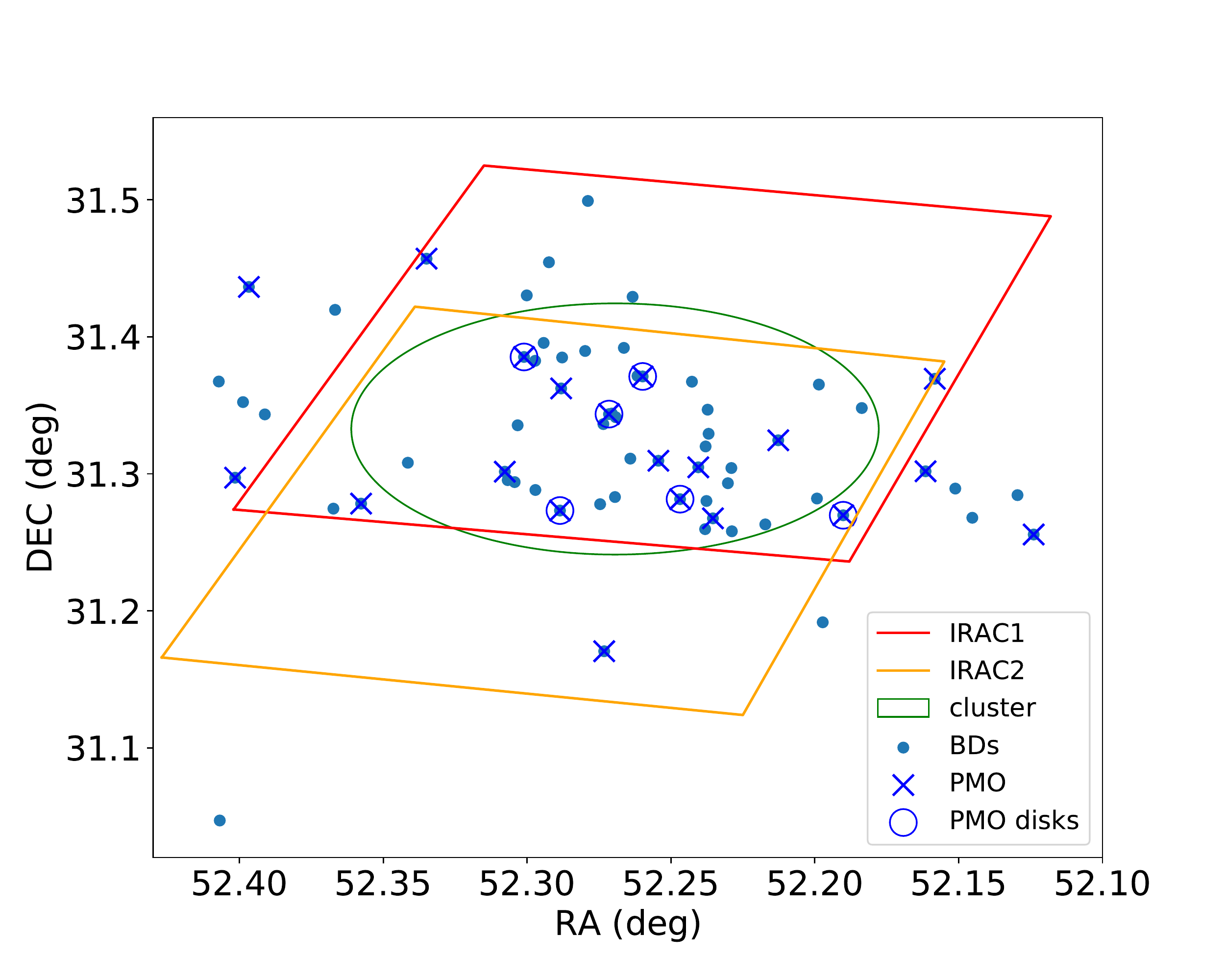}
\caption{{\bf Left:} Spectral type vs. K-band magnitudes for brown dwarfs in NGC1333. The dots mark all objects in this spectral type range from the list by \citet{luhman16}. The crosses mark the ones covered by our deep co-added Spitzer images. Spectral types are numerically coded, M6 is 6.0, L0 is 10.0. {\bf Right:} Spatial distribution of brown dwarfs (small dots), PMOs (crosses), and PMOs with disks (empty circles) in NGC1333. We also show the position and extent of the cluster as determined by \citet{gutermuth08}, as well as approximate coverage of the deep IRAC images produced here.
\label{sample}}
\end{figure*}

We characterise our sample in Figure \ref{sample}. In the left panel, we show spectral type vs. K-band magnitude for the whole brown dwarf catalogue, marking those covered by our deep images, and indicating approximate mass limits. As can be appreciated from this figure, our sample covers the entire substellar range, from the limit between stars and brown dwarfs to masses well below the Deuterium burning limit. In particular, it includes all objects with spectral types L0 or later found in this cluster thus far. In the right panel, we plot the spatial distributions of the brown dwarfs and PMOs (marking those with disks, see Section \ref{disc}). The figure demonstrates that our images cover the embedded cluster as well as surrounding areas.

\subsection{Photometry}

We carried out aperture photometry for the objects in our catalogue using tools from {\tt photutils} \citep{bradley21}. We chose a constant aperture with a radius of 5\,pixels. We note that the native pixel size of IRAC is 1.2", but the mosaics used here have 0.6" per pixel. The FWHM of IRAC during the Warm Mission is 2.0", according to the IRAC Instrument Handbook. An aperture radius of 5 pixels corresponds to a aperture diameter of 6", or approximately 3 times the FWHM, and should capture $>90$\% of the flux. For local background estimates, we measured the median flux in an annulus with inner and outer radius of 7 and 10\,pixels centered on the object coordinates. 

The magnitudes were calibrated by direct comparison with the JKS catalogue mentioned above. The offset between the JKS IRAC magnitudes, which originally come from the C2D source list \citep{evans09}, and our instrumental measurements is constant for a wide range of magnitudes, with standard errors of 0.03 and 0.04\,mag (depending on channel). Thus, simply adding an offset shifts our instrumental magnitudes into the standard system and makes them comparable with the literature.

Photometric errors were estimated by adding in quadrature the Poissonian errors in the source flux, the standard error for the background times the number of pixels in the aperture, and the error in the calibration. The median error, which is dominated by the calibration uncertainty, is 0.04\,mag for both channels.

All sources were checked in the stacked images. In particular, we verified that the background determined by the photometry routine is adequate and that the aperture is well centred on the source. As indicated earlier, some objects are located in area of highly variable background, which may affect the photometry.

Finally, we calculate the $K-IRAC1$ and $K-IRAC2$ color using the K-band values listed in \citet{luhman16}. We dereddened these colours to zero extinction using the published $A_J$ extinction values for the sample, again from \citet{luhman16}. Those values have been determined by comparing optical and near-infrared colours to expected photospheric values. For a subset of our sample, \citet{scholz12a} determined extinctions, using photometry and spectroscopy. Comparing with the Luhman values, there is good agreement, with a typical uncertainty $<0.5$\,mag in $A_J$.

For the extinction correction, we convert from $A_J$ to $A_K$ using a standard extinction law ($A_K = 0.382 A_J$, \citet{mathis90}). For the IRAC wavelengths, we adopt $A_{IRAC1} = 0.6 A_K$ and $A_{IRAC2} = 0.5 A_K$. These extinction relations are in line with recently published work for star forming regions, see \citet{chapman09}. With very few exceptions, the brown dwarfs in our sample have $A_J$ below 5. Thus, the exact choice of the extinction law within plausible values does not change the outcomes in any significant way. The dereddened colors vs. spectral types are plotted in Figure \ref{kirac}, as a way to identify objects with infrared excess emission and thus disks. These figures are the foundation for the discussion in Sect. \ref{disc}.

For completeness, we also did full-field photometry using the SExtractor \citep{bertin96} for the stacked image and an individual image in both channels. Comparing the peak of the histogram for the magnitudes, we find that the stacked images are deeper by 2.4 and 2.6\,mag, for IRAC1 and IRAC2 respectively. This result confirms the face-value expectation -- stacking 70 images of similar quality should result in $\sqrt{70}=8.3$ better signal-to-noise, or 2.3 magnitudes greater depth.

\section{Results and discussion}
\label{disc}

\subsection{Photospheric colours}

To determine which objects have excess emission in their infrared colours we first need to establish the photospheric levels in the same colours, as a function of spectral type. We prepared a sample of stars and brown dwarfs with spectral types from M0 to L1, all without disks (and thus without infrared excess), that are members of nearby star-forming regions ($<10$\,Myr). In this exercise, we are building on the work in \citet{almendros22}. The $K-IRAC1$ and $K-IRAC2$ colours for this sample show a gradual increase from values of 0.1 for early M dwarfs to around 1.0 for early L dwarfs. We fit these colours for spectral types M6 or later with a linear slope, which describes the trend seen in the data well. 
\begin{equation}
    K-IRAC1=0.1007\times SpT - 0.2847 
\end{equation}
\begin{equation}
    K-IRAC2=0.1179\times SpT - 0.3281
\end{equation}
Here, the spectral type ($SpT$) is 6.0 for M6 and 11.0 for L1. We tested these relations by doing the same procedure on only a subset of the diskless objects, and do not find a significant change. In particular, we limited the fit for objects with ages $<5$\,Myr and the fit remains very similar. For reference, the given relations are consistent with the ones published by \citet{luhman22} for ages $<20$\,Myr.

\begin{figure*}[t]
\center
\includegraphics[width=1.0\columnwidth]{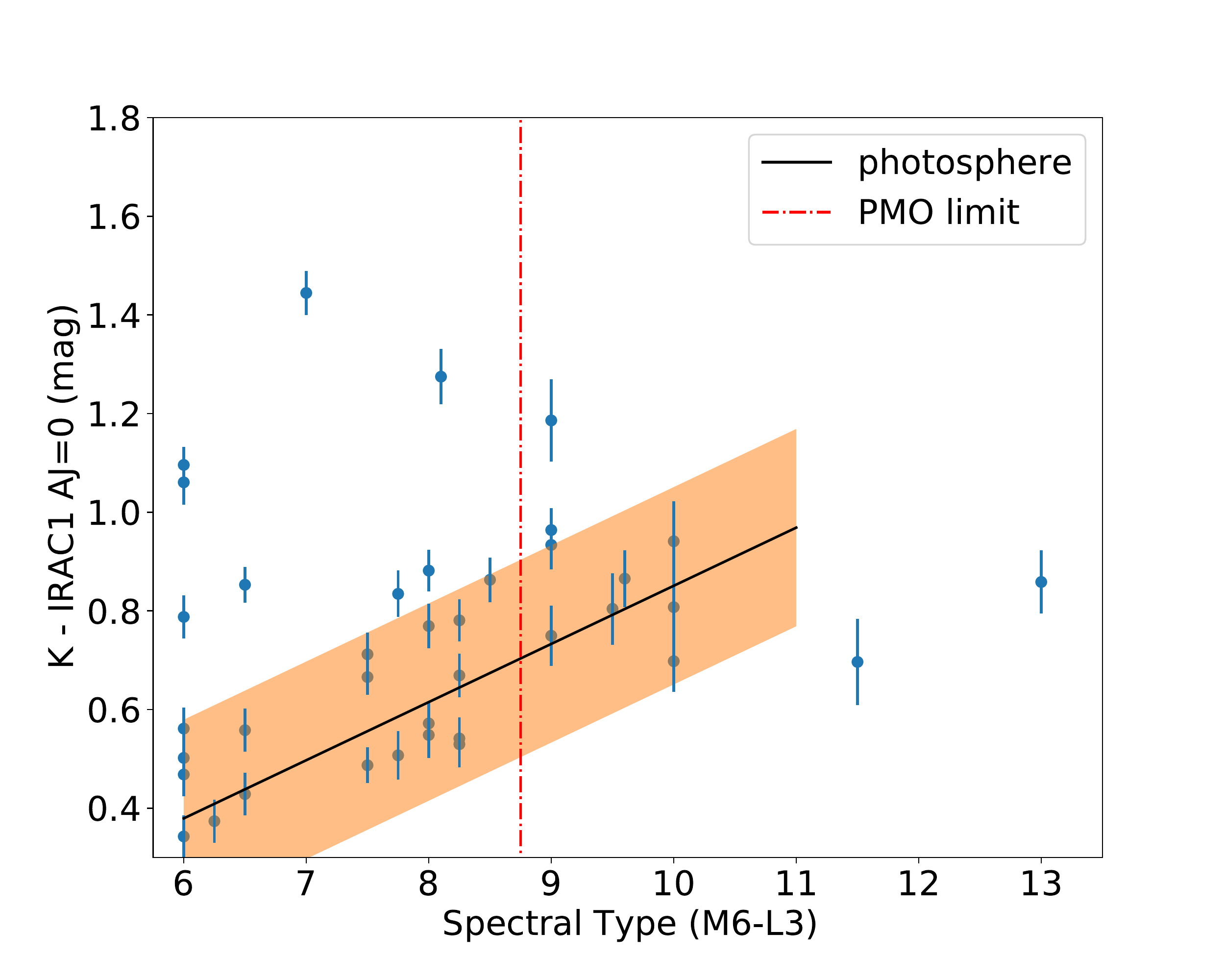}
\includegraphics[width=1.0\columnwidth]{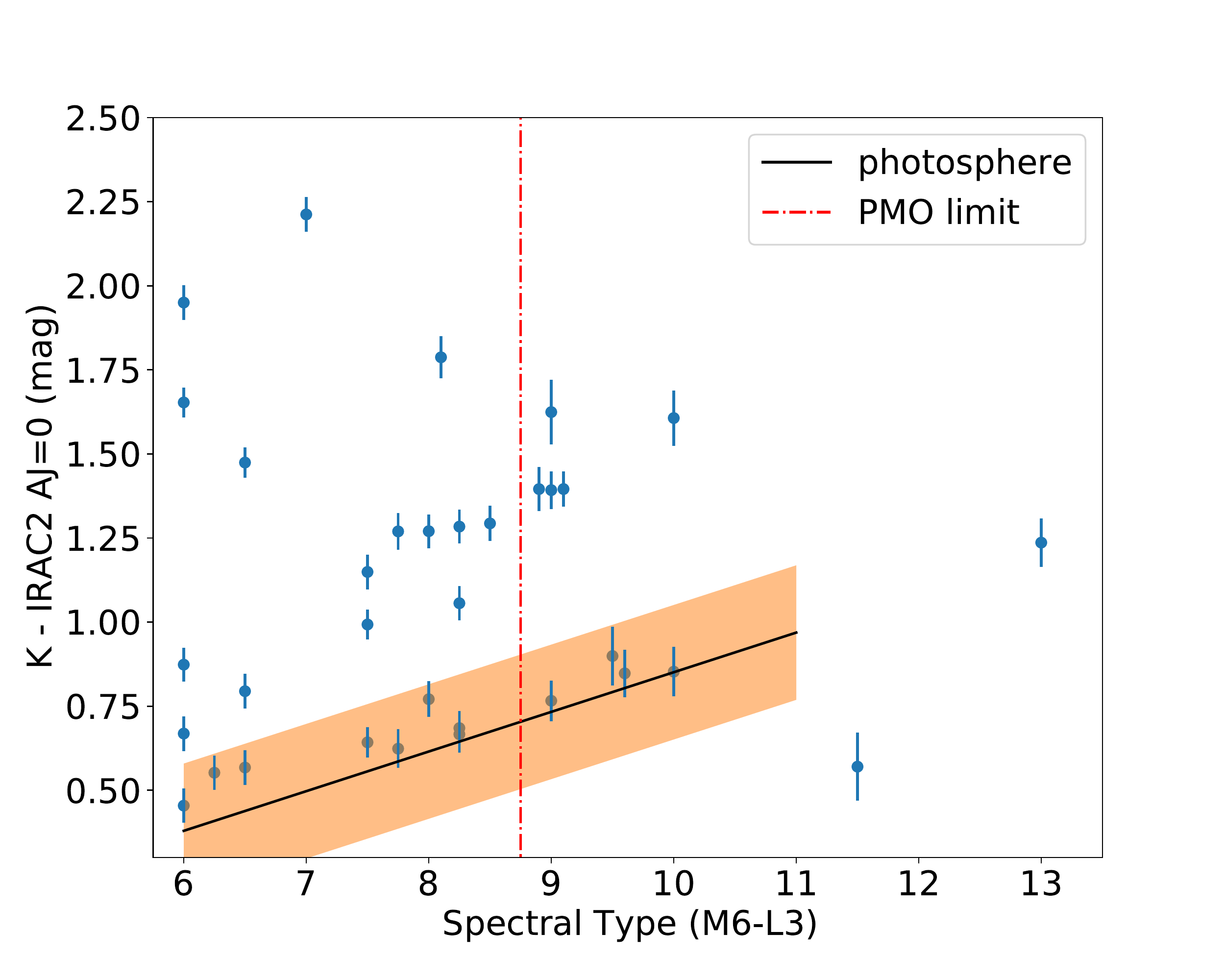}
\caption{Infrared colours of brown dwarfs in NGC1333 vs. spectral type, after dereddening the colours to $A_J=0$. The spectral type range is M6 to L3. In the right panel, three objects at spectral type M9 have very similar colours and have been plotted slightly offset in spectral type for clarity. The estimated photospheric colour is shown with a solid black line, with the typical scatter shown as coloured area (see text for details). The dash-dotted line is the chosen demarcation for our sample; all objects plotted on the right side of it can be described as planetary-mass objects (PMOs).
\label{kirac}}
\end{figure*}

Our photospheric colours are plotted in Figure \ref{kirac} as solid lines. We also show in the figure the typical scatter of diskless objects around this fit, as coloured area. Objects without infrared excess in NGC1333 are expected to be found in that area around the solid line in Figure \ref{kirac}. 

\subsection{Infrared excess and disk fraction}
\label{df}

In Table \ref{tab:pmo} we list the newly derived infrared photometry and colours (before extinction correction) for the 15 PMOs (spectral types M9 or later): 14 have measurements in IRAC1, 13 in IRAC2. Three objects with reported magnitudes are in areas of highly variable background, and after careful examination we deem those measurements to be unreliable. As can be seen in Figure \ref{kirac}, eight objects with good measurements have a dereddened $K-IRAC1$ colour consistent with photospheres, six do in $K-IRAC2$. Altogether seven sources out of 12 with robust measurements in at least one band show no evidence for infrared excess.

The remaining five objects are found to have infrared excess and are thus likely to be PMOs with disks. Four of these are at spectral type M9, three of them with excess in both bands (one with marginal excess in IRAC1). The fifth, with spectral type L0, has no excess in IRAC1, but a clear excess in IRAC2. Thus, the disk fraction from our sample is 5/12 or 42\%. Due to the small sample size, the plausible range for the disk fraction is 26-60\% (assuming 1$\sigma$ Binomial confidence intervals). Visual examination of the three objects in areas with highly variable background shows that one (SONYC-NGC1333-9) is bright in both IRAC bands, and thus also a good candidate for harboring a disk. The other two are faint and more likely to lack excess. Including those three would bring the disk fraction to 6/15, or 40\%.  We note that with one exception all disks in the PMO regime are found for brown dwarfs classified as M9, at the adopted upper limit for our PMO sample. For L0 or later objects, the disk fraction is only 1/5 or 20\%, 1/6 if we add the one with background-affected measurement.

Our analysis includes the lowest mass objects identified thus far in this cluster, in particular, the 6 objects with spectral type L0 or later (corresponding to masses of $\sim 0.01\,M_{\odot}$ or lower). We determine the IR excess for five out of six, with the one exception mentioned above. It is likely that the current optical/near-infrared surveys are not complete for those spectral types. Therefore, the disk fractions determined here may be affected by incompleteness at the lowest masses. If disk-bearing objects are on average deeper embedded than those without, thus have more extinction and are fainter, it is conceivable that we are underestimating the disk fraction for $M<0.01\,M_{\odot}$. This, in addition to small number statistics, may contribute to the the low fraction of disks in the L0 or later domain. Deeper observations will be needed to clarify.

In Figure \ref{sample}, right panel, we show the spatial distribution of the PMOs and PMOs with disks relative to the cluster extent and the overall brown dwarf sample. There is no obvious spatial bias in those samples, they are all concentrated on the cluster itself. One of the disk-bearing PMOs is located just outside the nominal cluster extent, but the same is true for several PMOs without disks.

Among the brown dwarfs with valid measurements and spectral types M6 to $<$M9 in our images, 8/27 show clear infrared excess in IRAC1, 15/23 in IRAC2. This corresponds to a disk fraction of $30\pm 10$\% for IRAC1 and $65\pm12$\% for IRAC2. Thus, based on our measurements, the disk fraction among PMOs is consistent within the statistical uncertainties with the value for more massive brown dwarfs. We only quote these numbers for completeness: In contrast to the faint PMOs, our deep images do not provide a significant benefit for the brighter M6 to $<$M9 objects. 

\subsection{Comparison with the literature}

In Table \ref{tab:pmo} we also included the previous classification of the infrared excess based on \citet{luhman16}, in the column labelled 'IR'. This previous paper uses mid-infrared excess measured from Spitzer images to identify circumstellar disks. Of our list of 15, 12 have that information listed in \citet{luhman16}, and for all 12 our classification is in agreement with the one in the literature. We also classify three objects for the first time here. One of them likely has a disk: at spectral type L0, it is currently the latest type object with a disk in NGC1333. We also classify the three objects with spectral types later than L0 as diskless, two of them for the first time.

\citet{luhman16} also derived disk fractions for subsamples divided by spectral type. They report 11/21 or 52\% for spectral types later than M8 (a slightly different subset than our sample of PMOs), 20/33 or 61\% for M6-M8, 28/48 or 58\% for M3.75-M5.75 and 27/39 or 69\% for K6 to M3.5. Within the statistical errorbars, these values are all consistent with each other, and with our own measurements of the disk fractions, given above. Our measurement is also consistent with the disk fraction of $66\pm8$\% for objects with spectral type of M5-M9 in NGC1333, reported by \citet{scholz12a}. For completeness, for a sample of 79 Gaia-selected low-mass stars in NGC1333, \citet{pavlidou22} measure a disk fraction of $67\pm13$\% (for the mass range 0.1 to 1.0\,$M_{\odot}$, approximately).

We note here that the `disk fraction' in the quoted papers and in our study is defined as the number of Class II objects, divided over the sum of Class II and Class III. These numbers are all based on samples from optical/near-infrared surveys, and should not be compared directly with the higher disk fractions derived from mid-infrared surveys \citep{gutermuth08,joergensen06}. One further caveat: Disk surveys based on photometry at 3-8$\,\mu m$ can by their nature not find disks with large inner holes where the disk only causes excess emission at longer wavelength. All studies mentioned above share this bias.

\begin{figure}[t]
\center
\includegraphics[width=0.9\columnwidth]{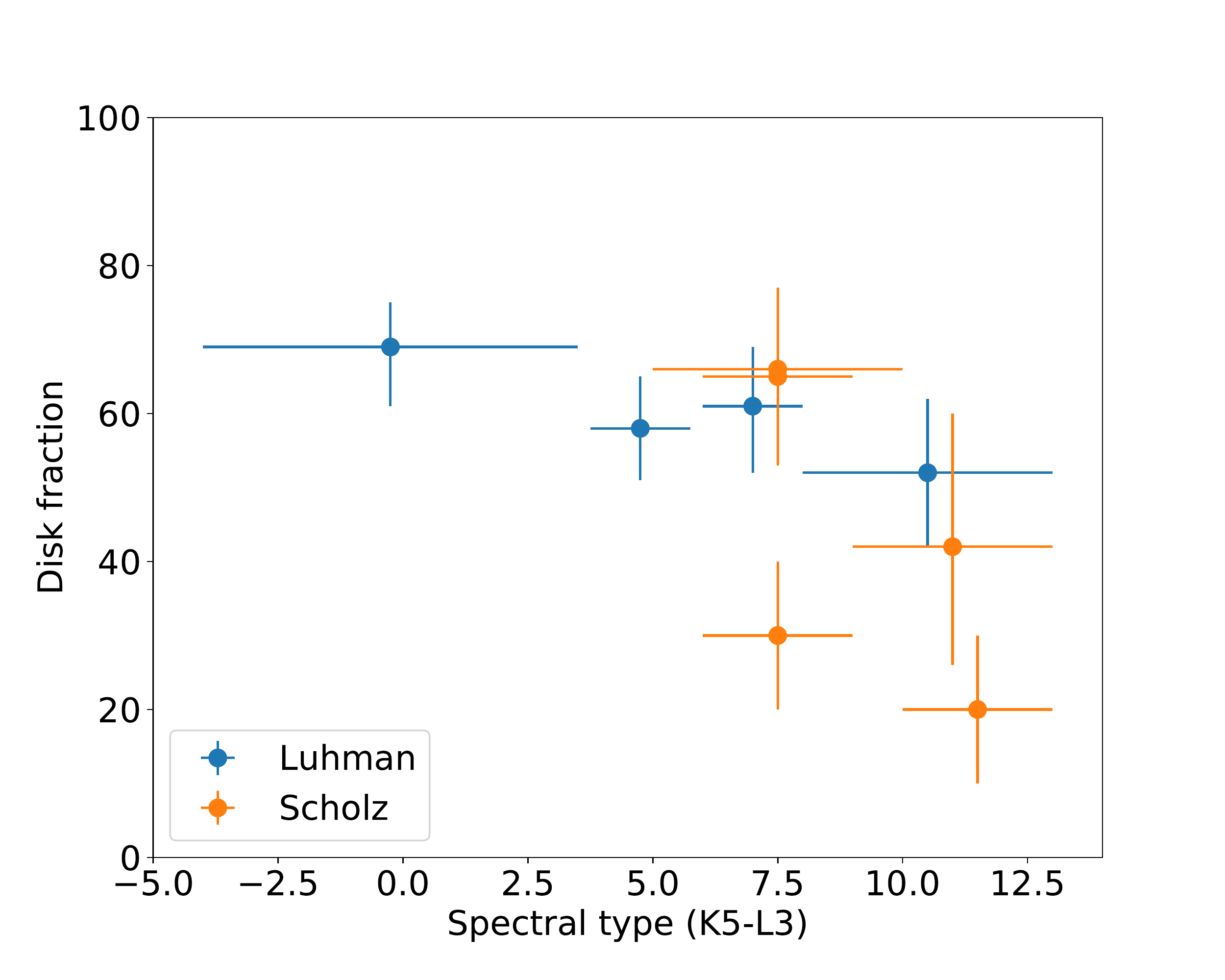}
\caption{Disk fraction for samples of stars and brown dwarfs with known spectral type, from \citet{luhman16} (blue symbols), \citet{scholz12a} and this paper (orange symbols). Spectral types are numerically coded, M0 is 0.0, L0 is 10.0.
\label{diskfraction}}
\end{figure}

In Figure \ref{diskfraction} we provide a summary of the currently available disk fraction measurements for low-mass stars, brown dwarfs, and PMOs in NGC1333, including only objects with known spectral type. As can be appreciated from this figure, there is no significant trend in the disk fraction from late K to early L spectral type. In that regard, our new photometry confirms results from previous studies of the disk population for this cluster. Finding a disk fraction largely independent of mass for low-mass stars and brown dwarfs is also in line with what has been found for other star-forming regions with ages between 1 and 5\,Myr, including IC348 \citep{luhman16}, $\sigma$\,Orionis \citep{scholz08}, and Chamaeleon-I \citep{luhman08a}. As a sidenote, the brown dwarf disk fraction derived in the current paper from IRAC1 excess (30\%) seems to be an underestimate compared with the literature and should be treated with caution.

Figure \ref{diskfraction} also illustrates that there may be a decline in the disk fraction when comparing L-type PMOs to K-M stars and brown dwarfs. As noted above, among the six cluster members in NGC1333 with spectral type L0 or later only one seems to have a disk, based on our measurements, translating to a disk fraction of 20\%. Given the small sample size and possible bias, as mentioned above, it is premature to conclude that this is indeed a drop-off in the disk fraction. If confirmed, it would constitute a very curious result, with possible implications for the formation process. Finding more L-type objects in this cluster would immediately give us a better handle on this issue.

We note that for somewhat older star-forming regions, like Upper Scorpius and the TW Hydrae Association, with ages $\sim$8-10\,Myr, there is some evidence in the literature for an increased disk fraction for brown dwarfs compared to low-mass stars \citep{luhman12,cook17,riaz08}. For those regions, PMOs do not seem to stand out from more massive brown dwarfs in terms of their disk fraction -- but the small sample size hampers a more definitive assessment of the longevity of disks around PMOs.

Generally speaking, the evidence is strong that free-floating PMOs can harbour disks, and therefore, have the potential to form their own (miniature) planetary systems. In terms of the overall mass and scale, these systems would be more comparable to Jupiter's Galilean moons than to the solar system. 

\subsection{The lowest mass free-floating objects with disks}

Our work in NGC1333 reported in this paper establishes that at least five PMOs in NGC1333 show evidence for the presence of a circum-sub-stellar disk. The lowest-mass among these is an object at spectral type L0 (not previously identified as disk-bearing), corresponding to a temperature of 2200\,K, which for an age of 1-2\,Myr would result in an estimated mass of $\sim$0.01\,$M_{\odot}$ \citep{baraffe15}. The three confirmed sources with later spectral types, and thus presumably lower masses, do not have detectable infrared excess and are classified as diskless.

Similar studies have been carried out in other star-forming regions. Taken together, there is now a small sample of young PMOs with securely identified disks. This includes OTS44 \citep{joergens13}, the only one  with an ALMA detection at far-infrared/submm wavelengths \citep{bayo17}. For young PMOs, definitively detecting an infrared excess due to a disk is challenging work, for two reasons. One, as can be appreciated from Figure \ref{kirac}, the photospheric near/mid-infrared colours increase steadily for young late M and L dwarfs. With the typical uncertainties in the spectral typing of at least $\pm1$ subtype, the photospheric level for a given source is uncertain. Second, these are faint sources, and thus, the near/mid-infrared magnitudes come with considerable errors. Both difficulties worsen towards later spectral types and lower masses. They can be overcome with deeper mid-infrared images (the motivation for this study), improved spectroscopic classification, or photometry at wavelengths where the photospheric flux is negligible compared to the disk excess ($10\,\mu m$ or beyond).

With all that in mind, we summarize the results for regions with sufficient survey depth.

In $\sigma$\,Orionis, two planetary-mass brown dwarfs with spectral types (SOri\,60 and SOri\,71) have consistently been found to host disks at wavelengths up to 8$\,\mu m$ \citep{scholz08,luhman08b}. These two have spectral types of L2 and L0 respectively. This cluster is slightly older than NGC1333; for its age of 3-5\,Myr, these spectral types would correspond to $\sim$0.01\,$M_{\odot}$ or above. There are some PMOs with later spectral types and claimed excess emission (e.g., SOri\,65, 66, 70), but the mid-infrared data are inconclusive \citep{scholz08,luhman08b}. In the case of SOri\,70, it is also not clear yet if it is in fact a young member of the cluster and not a foreground brown dwarf \citep{zapatero17}.

In Chamaeleon-I, OTS44 (M9.5) is a secure detection. In addition, \citet{luhman08a} found Cha 1107−7626, an L0 object with a disk. \citet{esplin17a} report three more objects with M9-L2/L3 spectral types that may have excess emission, but caution against interpreting it as evidence for a disk. Given the uncertainty in their spectral type, more work needs to be done before assigning masses or infrared excess to those.

In Taurus, the lowest mass objects with safely detected disks have late M spectral types, according to the recent survey by \citet{esplin19}. Some L0-L1 sources also may have infrared excess. In addition, \citet{esplin19} find one very low mass source with an L3 spectral type, which would have a mass well below 0.01\,$M_{\odot}$, with possible excess emission at 4.5$\,\mu m$. Again, this is not a robust detection yet.

$\rho$\,Ophiuchi has spectroscopically confirmed brown dwarfs with disks at spectral type M9-L1 \citep{testi02,jayawardhana06,alves12}, which again would correspond to a mass of $\sim$0.01\,$M_{\odot}$. CFHTWIR-Oph-33 has a spectral type of L3-L4 and may have excess emission, but the mid-infrared data are inconclusive in that regard \citep{alves12,esplin20}. Some other very low mass objects discussed in these papers may have infrared excess, but also very uncertain spectral type (e.g., CFHT-WIR-Oph-58, with a range from M8.5 to L3, see \citet{almendros22}).

Finally, the older Upper Scorpius association has a large population of brown dwarfs. Two with a spectral type of L3.5 are reported to have accretion and infrared excess \citep{lodieu18}. For ages of 5-10\,Myr, this would correspond to 0.01\,$M_{\odot}$ or above. To our knowledge, no lower mass objects with disks have been found here.

In summary, at this stage, the low-mass limit for objects with a {\it safe} disk detection, across all star forming regions, is $\sim 0.01\,M_{\odot}$ (or $\sim 10\,M_{\mathrm{Jup}}$). The surveys have found perhaps two dozen objects in total that may have masses below that limit (depending on how the mass is estimated) in all of the very young regions mentioned above. For the majority of them we can already firmly rule out the presence of disks -- the lowest mass objects in NGC1333 discussed in this paper belong in that category. For about ten of these PMOs with putative masses $<0.01\,M_{\odot}$, the currently available data are inconclusive and insufficient for proper characterisation. Therefore, it is certainly too early to rule out the presence of disks around objects with masses below $0.01\,M_{\odot}$. Having said that, given that the disk fraction among brown dwarfs is 30-60\% for the young regions mentioned above, the lack of confirmed disks below the $0.01\,M_{\odot}$ limit is perhaps already worth noting, especially if taken together with our new measurements for NGC1333, as illustrated in Figure \ref{diskfraction}. If the disk fraction were similar for objects with masses $<0.01\,M_{\odot}$, essentially all objects with tentative detections of infrared excess should have disks. Perhaps we are beginning to see a decline of disk fractions in the planetary-mass domain.

Sensitive mid-infrared observations of known PMOs, and deep surveys for new objects in this mass domain are needed to verify this suspicion. JWST is primed to help with both of these tasks. In NGC1333 specifically, a Guaranteed Time program with NIRISS/WFSS is dedicated to a deep spectroscopic survey, and designed to find PMOs in this cluster down to masses of 1$\,M{\mathrm{Jup}}$ (program ID 1202, see \citet{willott22}). Also, the unprecedented sensitivity of the MIRI instrument will enable us to probe directly for infrared excess in the lowest mass free-floating objects known, at wavelengths exceeding those of Spitzer/IRAC, allowing for unambiguous disk detections.

The presence of disks is discussed in the literature as one possible signature to distinguish between a star-like and a planet-like formation scenario \citep{luhman12b,testi16}. In particular, if an object gets ejected early in the evolution from a forming planetary system, either by planet-planet scattering or encounters with other stars, it can be expected that its disk, if present, will be affected. Simulations on this issue, however, remain sparse or non-existent. For the case of planet-planet-scattering, \citet{bowler11} demonstrate that the disruption of a circumplanetary disk should be common. 
In this context, disks detected with ALMA around young giant planets on wide orbits are an interesting comparison. These planetary-mass companions may have very compact disks \citep{wu17}, but it is unclear whether or not their masses are lower than expected from the standard relation between disk mass and stellar mass \citep{wu20}.\footnote{In the simulations by \cite{stamatellos15} objects formed in disks around stars are found to harbour their own disks with substantial masses, but the authors do not model the effect an ejection might have on those disks} 

Given the discussed arguments, it is plausible to assume that free-floating planets that have been ejected from a young planetary system would not host massive, long-lasting disks. If future observations confirm the absence of disks for objects with masses below $0.01\,M_{\odot}$, it may indicate that this is the low mass limit for objects to form like stars, physically set by the opacity limit for fragmentation.

\section{Summary}
\label{sum}

We present new Spitzer/IRAC photometry of brown dwarfs with planetary masses in the young star-forming cluster NGC1333. To improve on previous findings, we stack 70 images taken as a time series campaign during Spitzer's Warm Mission in 2011, at wavelengths of 3.6 and 4.5$\,\mu m$. Our deep images cover 50 brown dwarfs, 15 of which have spectral types of M9 or later, corresponding to masses near or below the Deuterium-burning limit (i.e., `planetary-mass objects', PMOs). Out of 12 PMOs with good measurements, five show clear infrared emission in excess of the estimated photospheric level. Taken together with previous work, our results confirm that the disk fraction does not change significantly between very low mass stars, brown dwarfs, and PMOs. Thus, free-floating objects with masses comparable to those of giant planets have the potential to form their own miniature planetary systems. However, we note that among the six lowest mass objects in NGC1333, with spectral types of L0 or later, only 1 has a clear disk detection (no. 12 in Table \ref{tab:pmo}), which may be a sign of a drop of in disk fraction at ultra-low masses, possibly suggesting that objects at these masses are primarily ejected planets, rather than formed from core collapse. We survey the literature for studies on disks around PMOs and find that the lowest mass objects with a firmly detected disks have masses around 0.01$\,M_{\odot}$. Some objects at later spectral types have tentative detections of infrared excess, but whether or not this is significant and due to the presence of a disk is still uncertain. Future observations with JWST will undoubtedly probe the presence of disks around lower mass objects, thus shedding light on the formation mechanism of free-floating PMOs. 

\begin{table*}[t]
\centering
\caption{Infrared photometry for planetary mass objects in NGC1333. Errors in the IRAC magnitudes 
are 0.03-0.09\,mag. \label{tab:pmo}}
\begin{tabular}{ccccccccccl}
\noalign{\smallskip}
\tableline
\noalign{\smallskip}
no & RA (deg) & DEC (deg) & SpT$^1$ & IRAC1 & IRAC2 & K-IRAC1 & K-IRAC2 & IR$^1$ & disk$^2$ & first spectrum \\
\noalign{\smallskip}
\tableline
\noalign{\smallskip}  
 1 & 52.15829 & 31.36939 & L0     & 16.72 &           &   0.98 &         & No  & No  & \citet{luhman16} \\
 2 & 52.19013 & 31.26978 & M9    & 16.24 & 15.58 &   0.79 & 1.45 & Yes & Yes & \citet{luhman16} \\
 3 & 52.21271 & 31.32450 & M9.5 & 16.84 & 16.74 &   0.85 & 0.95 & No  & No  & \citet{luhman16} \\
 4 & 52.23542 & 31.26753 & M9.6 & 15.94 & 15.96 &   0.87 & 0.85 & No  & No  & \citet{scholz09} \\
 5 & 52.24050 & 31.30478 & L1.5  & 16.73 & 16.74 &   1.15 & 1.14 &       & No  & \citet{esplin17b} \\
 6 & 52.24683 & 31.28153 & M9    & 15.02 & 14.52 &   1.10 & 1.60 & Yes & Yes & \citet{luhman16} \\
 7 & 52.25438 & 31.30956 & L0     & 16.29 & 16.11 &   0.83 & 1.01 & No  & No  & \citet{luhman16} \\
 8  & 52.25983 & 31.37108 & M9   & 15.69 & 15.26 &   1.19 & 1.62 & Yes & Yes & \citet{luhman16}\\
 9  & 52.27154 & 31.34361 & M9   & 11.64 & 10.77 &           &         & Yes & Yes? & \citet{luhman16}\\
 10& 52.27321 & 31.17053 & M9   &           & 15.12 &           & 0.77 & No  & No  & \citet{scholz09} \\
 11& 52.28817 & 31.36233 & M9   & 10.84 & 10.65 &           &         & No  & No? & \citet{luhman16} \\
 12& 52.28858 & 31.27322 & L0    & 17.21 & 16.53 &   0.98 & 1.66 &       & Yes & \citet{luhman16} \\
 13& 52.30108 & 31.38531 & M9   & 13.77 & 13.23 &   1.39 & 1.93 & Yes & Yes & \citet{luhman16}\\
 14& 52.30775 & 31.30158 & L1    & 17.43 &           &           &         &       & No?  & \citet{scholz12b} \\
 15& 52.35771 & 31.27828 & L3    & 16.42 & 16.04 &   0.86 & 1.24 & No  & No  & \citet{scholz12a} \\
\noalign{\smallskip}
\tableline
\noalign{\smallskip}
\end{tabular}

$^1$ from \citet{luhman16}; $^2$ our classification, based on Figure \ref{kirac}
\end{table*}

\section{Acknowledgements}

We thank the anonymous referee for a constructive and concise referee report on this paper. This work is based on observations made with the Spitzer Space Telescope, which was operated by the Jet Propulsion Laboratory, California Institute of Technology under a contract with NASA. VA and KM acknowledge funding by the Science and Technology Foundation of Portugal (FCT), grants No. {\tt PTDC/FIS-AST/7002/2020},  {\tt UIDB/00099/2020} and {\tt SFRH/BD/143433/2019}. Constructive discussions with Dimitris Stamatellos, Ken Rice, Ian Bonnell, and James Wurster have aided the discussion in this paper. In order to meet institutional and research funder open access requirements, any accepted manuscript arising shall be open access under a Creative Commons Attribution (CC BY) reuse licence with zero embargo.

\end{document}